\documentclass[aps,prl,showpacs,twocolumn,amsmath,10pt,superscriptaddress,floatfix,nofootinbib]{revtex4-1}
\usepackage{epsfig,amssymb,amsfonts,bm}
\usepackage{array}
\usepackage[active]{srcltx}
\usepackage{color}
\usepackage[colorlinks=true,linkcolor=blue,citecolor=blue]{hyperref}
\newcolumntype{L}{>{$}l<{$}} 
\newcolumntype{C}{>{$}c<{$}} 

\newcommand{\be}{\begin{equation}}
\newcommand{\ee}{\end{equation}}
\newcommand{\bea}{\begin{eqnarray}}
\newcommand{\eea}{\end{eqnarray}}
\newcommand{\beas}{\begin{eqnarray*}}
\newcommand{\eeas}{\end{eqnarray*}}

\newcommand{\GeV}{\,\mbox{GeV}}

\newcommand{\itp}{\affiliation{CAS Key Laboratory of Theoretical Physics,
            Institute of Theoretical Physics,\\ Chinese Academy of Sciences,
            Beijing 100190, China}}
\newcommand{\bonn}{\affiliation{Helmholtz-Institut f\"ur Strahlen- und
             Kernphysik and Bethe Center for Theoretical Physics,\\
             Universit\"at Bonn,  D-53115 Bonn, Germany}}
 \newcommand{\JSC}{\affiliation{JARA-HPC, Jülich Supercomputing Center, Forschungszentrum Jülich, 54245 Jülich Germany}}
\newcommand{\fzj}{\affiliation{Institute for
           Advanced Simulation, 
           Forschungszentrum J\"ulich, D-52425 J\"ulich, Germany}}
\newcommand{\ucas}{\affiliation{School of Physical Sciences,
            University of Chinese Academy of Sciences,
            Beijing 100049, China}} 
\newcommand{\ikp}{\affiliation{Institut f\"ur Kernphysik and
           J\"ulich Center for Hadron Physics,\\ 
           Forschungszentrum J\"ulich, D-52425 J\"ulich, Germany}}

\begin{document}

\title{Confirmation of the existence of an exotic
state in the $\pi D$ system}

\author{Eric B. Gregory} \email{e.gregory@fz-juelich.de}\fzj\JSC
\author{Feng-Kun~Guo}\email{fkguo@itp.ac.cn}\itp\ucas
\author{Christoph~Hanhart} \email{c.hanhart@fz-juelich.de}\fzj\ikp
\author{Stefan Krieg} \email{s.krieg@fz-juelich.de}\fzj\JSC\bonn
 \author{Thomas Luu} \email{t.luu@fz-juelich.de} \fzj\ikp\bonn


\begin{abstract}
In recent years many candidates for states beyond the most simple realization of the quark model were found in various experiments around the world. However, so far no consensus exists on their structure, although there is strong evidence that at least some of those are dynamically generated from meson-meson interactions. In this Letter we provide an important missing piece from the theoretical side to prove that 
the lightest open charm strange and nonstrange scalars $D_{s0}^*(2317)$ and $D_0^*$ as well as their axial-vector partner states can all be understood as emerging from the interactions between 
Goldstone bosons stemming from the spontaneous breaking of chiral symmetry 
and the ground state charmed mesons.
For that purpose we exploit the flavor multiplet structure of the lightest open-charm positive-parity scalar states in an SU(3) symmetric lattice QCD simulation at large pion masses
to establish that there exists a bound state in the flavor-sextet representation, which cannot emerge for quark-antiquark states but appears naturally for four-quark configurations. Moreover, we find repulsion in the $[15]$ representation and thus no single-particle state in this representation exists, falsifying the expectation for tetraquark models. The findings establish the pattern predicted for the interactions of Goldstone bosons with $D$ mesons from chiral symmetry and the paradigm of the lowest-lying positive-parity charmed mesons as dynamically generated states.
\end{abstract}

\maketitle

{\it Introduction.}---Because of color confinement of quantum chromodynamics (QCD), understanding the spectrum of hadrons is one of the most challenging tasks in the study of the strong interaction. Since 2003, when the $B$-factories entered the hunt for
the hadronic spectrum,
many hadrons were observed with properties in conflict with the predictions from conventional quark models that identify mesons as $\bar qq$ states. 
For example, the $D_{s0}^*(2317)$~\cite{Aubert:2003fg} and $D_{s1}(2460)$~\cite{Besson:2003cp} are significantly
lighter than the predicted scalar and axial-vector $c\bar s$ states at $2.48\GeV$ and $2.55\GeV$, respectively~\cite{Godfrey:1985xj,Godfrey:2015dva,Ebert:2009ua}.
This led to the development of various models, including $D^{(\ast)}K$ hadronic molecules~\cite{Barnes:2003dj,vanBeveren:2003kd,Szczepaniak:2003vy,Kolomeitsev:2003ac,Chen:2004dy,Guo:2006fu,Guo:2006rp,Gamermann:2006nm}, tetraquark states~\cite{Cheng:2003kg,Maiani:2004vq}, and mixtures of $c\bar{q}$ with tetraquarks~\cite{Browder:2003fk,Dai:2006uz}. 

The corresponding charm-nonstrange states are known as $D_0^*(2300)$~\cite{Abe:2003zm,Link:2003bd}
and $D_1(2430)$~\cite{Abe:2003zm}.
Clearly their masses are significantly higher than expected from the SU(3) breaking pattern, since they are too close to those of the partner states
containing strangeness.
This puzzle and the fact that the masses
of the $D_{s0}^*(2317)$ and $D_{s1}(2460)$ are equidistant
to the $DK$ and $D^*K$ thresholds, respectively,
get naturally understood employing unitarized chiral perturbation theory (UChPT). 
For the singly heavy states this approach was pioneered in Ref.~\cite{Kolomeitsev:2003ac} --- for more recent developments see, e.g., Refs.~\cite{Albaladejo:2016lbb,Du:2017zvv,Guo:2017jvc}. This formalism
 allows one to calculate the nonperturbative dynamics of the scattering of the Goldstone bosons of the spontaneously broken chiral symmetry off the $D_{(s)}^{(*)}$ mesons in a controlled way. 
In Ref.~\cite{Liu:2012zya} a fit to lattice data for various Goldstone-boson--$D_{(s)}$-meson
scattering lengths (but not in the channel where
the $D_{s0}^*(2317)$ is) fixed the a priori unknown low-energy constants. The resulting amplitudes
not only reproduced the correct $D_{s0}^*(2317)$ mass,
but also predicted its pion mass dependence~\cite{Du:2017ttu} that agrees well with lattice results~\cite{Bali:2017pdv}. 
In addition, this study solved the mass hierarchy puzzle mentioned above by confirming that the structure known as
$D_0^*(2300)$ emerges from two poles, one lighter and one heavier, in line with other findings~\cite{Kolomeitsev:2003ac,Guo:2006fu,Guo:2009ct,Guo:2015dha,Albaladejo:2016lbb,Guo:2018tjx,Guo:2018kno,Guoa:2018dhm}. 
Their pole locations are found at 
$\left(2105^{+6}_{-8}-i\,102^{+10}_{-11}\right)\text{MeV}$ and $\left(2451^{+35}_{-26}-i\, 134^{+7}_{-8}\right)\text{MeV}$~\cite{Albaladejo:2016lbb,Du:2017zvv}, respectively. The SU(3) partner of the $D_{s0}^*(2317)$ is the lighter one, which restores the expected mass hierarchy. 
The heavier pole on the other hand is a member of SU(3) sextet, which is exotic from the conventional quark model.
Support for the presence of two poles comes from an analysis of the high-quality LHCb data on the decays $B^-\to D^+\pi^-\pi^-$~\cite{Aaij:2016fma}, $B_s^0\to\bar{D}^0K^-\pi^+$~\cite{Aaij:2014baa},
$B^0\to \bar{D}^0 \pi^-\pi^+$~\cite{Aaij:2015sqa}, $B^-\to D^+\pi^- K^-$~\cite{Aaij:2015vea}, and
$B^0\to\bar{D}^0\pi^- K^+$~\cite{Aaij:2015kqa} performed
in Refs.~\cite{Du:2017zvv,Du:2019oki,Du:2020pui}, as well as from the fact that their existence is consistent with the lattice energy levels~\cite{Liu:2012zya,Mohler:2013rwa,Lang:2014yfa,Moir:2016srx,Bali:2017pdv,Cheung:2020mql} for the relevant two-body scattering~\cite{Albaladejo:2016lbb,Albaladejo:2018mhb,Guo:2018tjx}. Furthermore, the predicted $D\bar K$ virtual state~\cite{Albaladejo:2016lbb} 
in the sextet and that the lightest $D_0^*$ should be significantly lighter than 2.3~GeV were also confirmed in lattice QCD (LQCD)~\cite{Moir:2016srx,Gayer:2021xzv}.
In addition, it is shown in Ref.~\cite{Du:2020pui} that the $D_0^*(2300)$ with resonance parameters listed in the Review of Particle Physics~\cite{Zyla:2020zbs} is in conflict with the LHCb data on $B^-\to D^+\pi^-\pi^-$~\cite{Aaij:2016fma}, contrary to the two $D_0^*$ states scenario.
This two-pole structure indeed emerges as a more general pattern in the hadron spectrum, see, e.g., Ref.~\cite{Meissner:2020khl}.

However, from experimental data alone the underlying flavor structure cannot be resolved model-independently. 
Thus, regardless of the successes of the molecular picture described above,
the controversy about the structure of the lightest
positive parity charmed states
is still not resolved. A direct measurement of the width
of the $D_{s0}^*(2317)$ will provide valuable additional evidence for or against a molecular nature, since the hadronic molecular component enhances the hadronic widths by an order of magnitude~\cite{Faessler:2007gv,Lutz:2007sk,Guo:2008gp,Liu:2012zya,Guo:2018kno} which could be measured at PANDA~\cite{Mertens:2012kpa}. Another important piece to complete the puzzle is 
to study the multiplet structure: Since flavor SU(3) is an approximate symmetry of QCD it leaves an imprint in the spectrum that can be organized in SU(3) multiplets. As outlined in the next section, the multiplet structure depends on the nature of the states. 
Accordingly, establishing the existence of the higher $D_0^*$ state in LQCD together with the multiplet it belongs to provides crucial information to pin down the nature of the lightest positive-parity charmed mesons and to understand the overall hadron spectrum from QCD. 

To establish the existence of a second $D_0^*$ state
as well as its underlying structure
unambiguously, we follow the strategy suggested in Ref.~\cite{Du:2017zvv}. It requires
lattice QCD simulations at an SU(3) flavor symmetric point  for masses of the light
 pseudoscalar-mesons mass in the range 600--700~MeV.
 The theory predicts at the corresponding unphysically high quark masses a bound or virtual state
 in the flavor sextet that cannot emerge in the
 traditional $c\bar u$ quark models.

{\it $SU(3)$ multiplets for different scenarios.}---The charm quark $c$ sits in an SU(3) singlet, up, down and strange quarks, below collectively called $q$, form a triplet and accordingly their anti-quarks, $\bar q$, an antitiplet. 
Thus, states with a $c\bar q$ structure sit exclusively in the flavor $[\bar 3]$ irreducible representation (irrep).

If we follow the standard assignment, compact tetraquarks are made of compact diquarks in the color-antitriplet representation~\cite{Maiani:2004vq}
although some works also discuss the color-sextet configuration~\cite{Dmitrasinovic:2012zz,Richard:2017vry}. Then,
the Pauli principle demands that a $[\bar q\bar q]$ diquark sits either in the strongly attractive flavor $[3]$ irrep with spin 0 or the less attractive $[\bar 6]$ with spin 1. If we couple this to the $[3]$ of the $[cq]$ diquark, 
we find the following flavor multiplets of states
\begin{eqnarray}
[3]\otimes[3] = [\bar 3]\oplus[6], \qquad
[\bar 6]\otimes[3] = [\bar 3]\oplus [15] \, ,
\end{eqnarray}
with the former being more strongly bound than the latter. 
Note that if indeed diquarks were compact constituents within tetraquarks,
the interactions among them would be analogous to the interactions within
regular mesons since both share the same color structure of their constituents.
Accordingly, in these tetraquark models 
all the above flavor multiplets should be manifest in the mass spectrum.
In particular, one should find attraction in all multiplets, keeping in mind
 that the two flavor $[\bar 3]$ representations can mix.
Note that in the tetraquark model of Ref.~\cite{Dmitrasinovic:2005gc} the non-strange
member of the [15] appears even lower in mass
than that of the [6] as a consequence of the 
action of the t'Hooft force, although the actual ordering depends very much on the details of the model employed~\cite{Dmitrasinovic:2012zz}.

Finally, for the hadronic molecular picture we need to scatter the octet ($[8]$) of Goldstone bosons off the 
flavor-antitriplet ($[\bar 3]$) of the charmed mesons to find
\begin{equation}
[8]\otimes[\bar 3]=[\bar 3]\oplus[6]\oplus[15] \, .
\end{equation}
To see in which of those one expects attraction,
one may employ chiral perturbation theory to leading order --- the
mass pattern that emerges from a proper unitarisation of this interaction~\cite{Kolomeitsev:2003ac,Guo:2006fu,Guo:2009ct}
already shows the features 
of the more refined treatments~\cite{Guo:2015dha,Albaladejo:2016lbb,Guo:2018tjx,Guo:2018kno,Guoa:2018dhm}. This study shows that the interaction in
the $[\bar 3]$ is most attractive, that
in the $[6]$ is less attractive and the one in the $[15]$ is repulsive~\cite{Hyodo:2006kg,Albaladejo:2016lbb}. 

Thus, to establish the multiquark structure of the lightest positive-parity charmed mesons it appears sufficient to show that in addition to the most strongly bound state there is a state in the $[6]$ representation of SU(3). To also disentangle hadronic molecules and tetraquarks it needs to be determined, whether the interaction in the $[15]$ is repulsive or attractive,
respectively.

{\it SU(3) interpolating operators.}---The general construction of interpolating operators for the $[6]$ and $[15]$ is done by first constructing the needed SU(3) flavor states within a tensor basis~\cite{Georgi:1982jb}.  Since the $c$ quark is in an SU(3) singlet, the states for the remaining degenerate light quarks can be constructed with appropriate projection onto the $[6]$ and $[15]$ irreps.  
\begin{table}
\center
\caption{The light quark content of the states in the $[6]$ representation and their associated quantum numbers.  $T^2_a$ is the Casimir operator, $I_z$ is the third component of isospin, and $Y$ is the hypercharge.\label{tab:6}}
\begin{ruledtabular}
\begin{tabular}{c|c|c|c|c}
state&components& $T_a^2$ & $I_z$ & $Y$\\
\hline\hline
1& $-|u\bar{u}\bar{d}\rangle\frac{1}{2}+|u\bar{d}\bar{u}\rangle\frac{1}{2}-|s\bar{d}\bar{s}\rangle\frac{1}{2}+|s\bar{s}\bar{d}\rangle\frac{1}{2}$ &$\frac{10}{3}$ &$+\frac{1}{2}$ & $-\frac{1}{3}$\\
2& $|d\bar{u}\bar{d}\rangle\frac{1}{2}-|d\bar{d}\bar{u}\rangle\frac{1}{2}-|s\bar{u}\bar{s}\rangle\frac{1}{2}+|s\bar{s}\bar{u}\rangle\frac{1}{2}$ &$\frac{10}{3}$ &$-\frac{1}{2}$ & $-\frac{1}{3}$\\
\hline
3& $|u\bar{u}\bar{s}\rangle\frac{1}{2}-|u\bar{s}\bar{u}\rangle\frac{1}{2}-|d\bar{d}\bar{s}\rangle\frac{1}{2}+|d\bar{s}\bar{d}\rangle\frac{1}{2}$ &$\frac{10}{3}$ &0 & $+\frac{2}{3}$\\
4& $|d\bar{s}\bar{u}\rangle\frac{1}{\sqrt{2}}-|d\bar{u}\bar{s}\rangle\frac{1}{\sqrt{2}}$ &$\frac{10}{3}$ &$-1$ & $+\frac{2}{3}$\\
5& $|u\bar{s}\bar{d}\rangle\frac{1}{\sqrt{2}}-|u\bar{d}\bar{s}\rangle\frac{1}{\sqrt{2}}$ &$\frac{10}{3}$ &$+1$ & $+\frac{2}{3}$\\
\hline
6& $|s\bar{d}\bar{u}\rangle\frac{1}{\sqrt{2}}-|s\bar{u}\bar{d}\rangle\frac{1}{\sqrt{2}}$ &$\frac{10}{3}$ &0 & $-\frac{4}{3}$\\
\end{tabular}
\end{ruledtabular}
\end{table}
For example, \autoref{tab:6} shows the states and their associated quantum numbers for the $[6]$ irrep.  A similar table can be given for the 15 states in the $[15]$ irrep.  These states are purely in flavor space, and must be coupled with spinors to ultimately construct interpolating operators.  To do this, we couple any of the states in \autoref{tab:6} with the singlet $c$ state and insert the relevant spinor structure $\Gamma$ at appropriate places.  For example, using the fifth state of \autoref{tab:6} and ignoring any overall minus sign, one has the following interpolating operator
\begin{multline}
O^5_{[6]}(x';x)=\frac{1}{\sqrt{2}}\left\{\left[\bar{s}(x')\Gamma c(x')\right]\left[\bar{d}(x)\Gamma u(x)\right]\right.\\-\left.\left[\bar{d}(x')\Gamma c(x')\right]\left[\bar{s}(x)\Gamma u(x)\right]\right\}\ .
\end{multline}
In our calculations we have $\Gamma=\gamma_5$.  To construct our correlators we must ultimately contract our interpolating operators $\langle O(y';y)\bar{O}(x;x)\rangle$  using Wick's theorem.  Here we assume all sources are located at $x$ and sink locations $y'$ and $y$ span the entire space-time volume (point-to-all \cite{Gattringer:2010zz}).  We note that the contractions are diagonal within the states of each irrep.  Furthermore, the degeneracy at the SU(3) symmetric point ensures that all contractions are \emph{identical} within each irrep. When comparing the contractions for the $[6]$ and $[15]$ states, we find that they differ by a relative minus sign in their exchange term, 
\begin{multline}\label{eqn:6 contraction}
\langle O^{i}_{[d]}(y';y)\bar{O}^{i}_{[d]}(x;x)\rangle = \\
\text{Tr}\left[\Gamma \gamma_5\mathcal{S}^\dag_{y';x}\gamma_5\Gamma\mathcal{S}^{}_{y';x}\right]
\text{Tr}\left[\Gamma \gamma_5\mathcal{S}^\dag_{y;x}\gamma_5\Gamma\mathcal{C}^{}_{y;x}\right]\\
\pm\text{Tr}\left[\Gamma \gamma_5\mathcal{S}^\dag_{y;x}\gamma_5\Gamma\mathcal{S}^{}_{y';x}\Gamma \gamma_5\mathcal{S}^\dag_{y';x}\gamma_5\Gamma\mathcal{C}^{}_{y;x}\right]\ ,
\end{multline}
where $\text{Tr}$ represents a trace over spin and color and the plus (minus) sign is for the $d=6(15)$ states. Here $\mathcal{S}$ and $\mathcal{C}$ are the light and charm quark propagators, respectively. Notice that the second term in the above equation with opposite signs in the [6] and [15] irreps is the leading contribution in the $1/N_c$ expansion, with $N_c$ the number of colors, while the first term is subleading~\cite{Guo:2013nja}, indicating opposite signs for the corresponding interactions.

In our LQCD calculations we use a Sobol sequence to sample the source locations $x$.  At the sink side we project each hadron to zero momentum by performing $\frac{1}{L^6}\sum_{\vec{y}',\vec{y}}$, obtaining excellent overlap with the ground state energy.  All our results presented in the next section were obtained in this manner.  We note that we have also projected the combined two-hadron system to zero momentum 
by setting $\vec{y}'=\vec{y}$ and performing $\frac{1}{L^3}\sum_{\vec{y}}$.  In this case the correlators had significant contamination from thermal and excited states, and though our extracted energies were consistent with the former projection, because of their large uncertainties we did not include these results in our analysis.

{\it Lattice parameter tuning.}---We performed a simulation using  $N_f=3+1$ flavors of  clover-improved Wilson~\cite{Sheikholeslami:1985ij} fermions with six iterations of stout link smearing~\cite{Morningstar:2003gk}. We use the CHROMA~\cite{Edwards:2004sx} QCD software system, supported by either the QPhiX~\cite{osti_1324431} or QUDA~\cite{Clark:2009wm} inverter libraries, depending on hardware architecture.
We aimed to simulate at a target point in parameter space where the charm quark mass $m_c$ is physical and tune the light quark mass  $m_q$ such that the $q\overline{q}$ pion mass falls in the range $600~{\rm MeV} <M_\pi < 700$~MeV. 

On tuning ensembles, we calculate  meson masses  with $q\overline{q}$ and $c\overline{c}$ and associate the $q\overline{q}$ pseudoscalar with the pion, and the $c\overline{c}$ pseudoscalar and vector mesons with the $\eta_c$ and $J/\psi$, respectively. We do not include disconnected diagrams in the charmonium meson correlators. This induces systematic errors in both the quark mass tuning and lattice spacing calibration. Neither are relevant to the conclusions of this Letter.
 
\begin{figure}
    \begin{center}
    \includegraphics[width=\columnwidth]{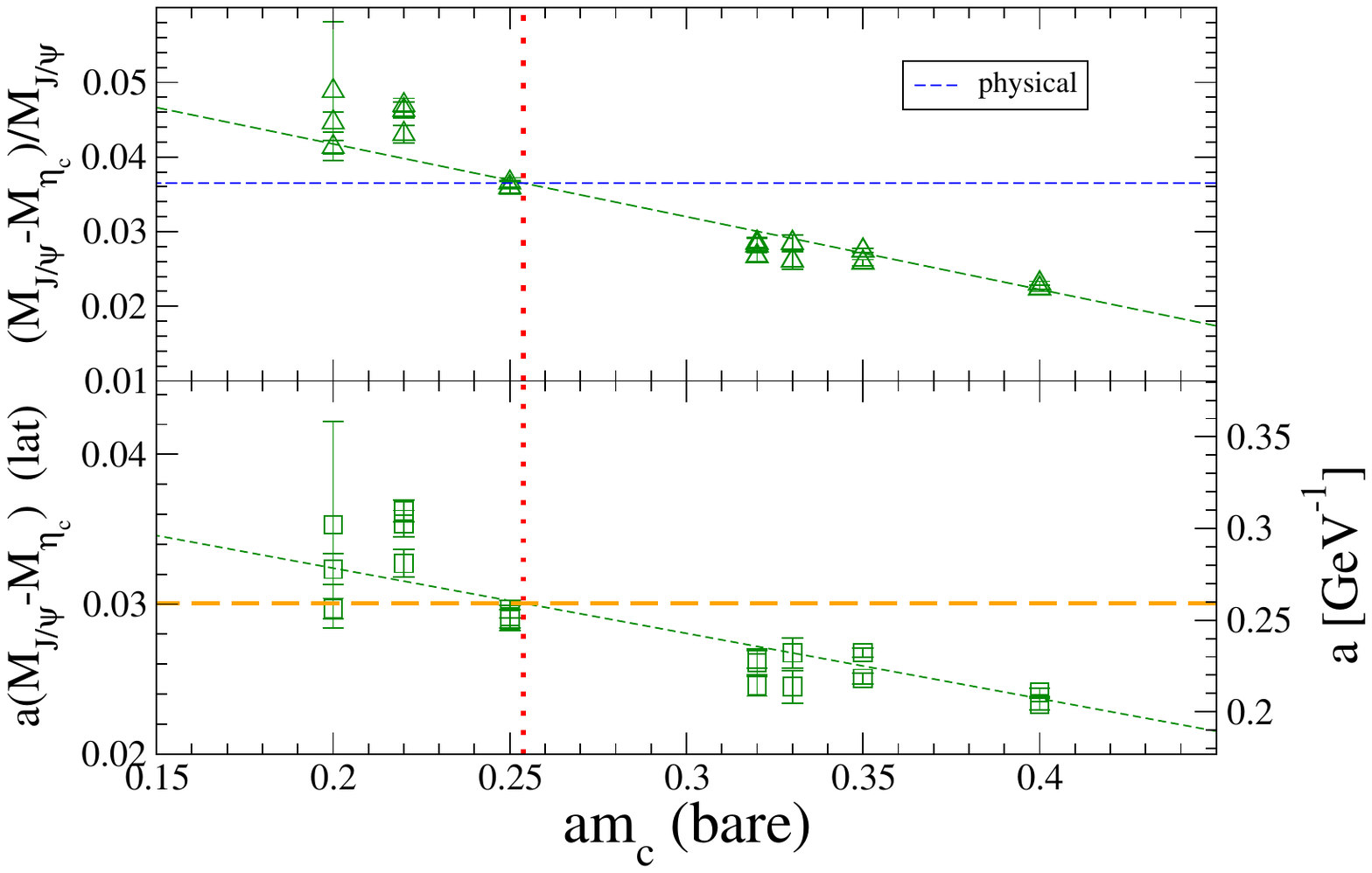}
    \end{center}
    \caption{(Top) Tuning $m_c$. We find the value of $am_c$ where the linear fit to measured values of the charmonium splitting ratio (green dashed line) crosses the physical value $R=0.0365$ (blue dashed line).\\
    (Bottom) Determining the lattice scale $a$. Where the linear fit to lattice values of the splitting $a(M_{J/\psi}-M_{\eta_c})$ intersects this same value of $am_c$, we use this splitting to determine the lattice spacing $a$ in GeV$^{-1}$ (horizontal orange dashed line to the right axis). 
    \label{fig:combi_mc_tune}}
\end{figure}

We find the target in parameter space through a tuning procedure of several steps.
First, we tune $m_c$ until the dimensionless splitting ratio 
\begin{equation}\label{eq:split_rat}
    R_{c\overline{c}}\equiv \frac{M_{J/\psi}-M_{\eta_c}}{M_{J/\psi}}
\end{equation}
is at its physical value 0.0365. For a given $\beta$ value $R_{c\overline{c}}$ varies roughly linearly with $m_c$, as shown in Fig.~\ref{fig:combi_mc_tune} (top). 
Once we determine the target $m_c$ for each $\beta$, we use the lattice value of the splitting $aM_{J/\psi}-aM_{\eta_c}$ to establish the lattice spacing:
\begin{equation}\label{eq:lat_spac}
a=\frac{\left(aM_{J/\psi}-aM_{\eta_c}\right)_{\rm latt}}{\left(M_{J/\psi}-M_{\eta_c}\right)_{\rm phys}} = \frac{\left(aM_{J/\psi}-aM_{\eta_c}\right)_{\rm latt}}{113 ~{\rm MeV}}.
\end{equation}
For $\beta=3.6$ we determine $a=0.27(2)_{\rm stat}(2)_{\rm sys}$ GeV$^{-1}$. The neglect of disconnected diagrams induces a systematic overestimate of $\left(aM_{J/\psi}-aM_{\eta_c}\right)_{\rm latt}$~\cite{Hatton:2020qhk}, and hence a proportional error on $a$.

The final step is to tune the light quark mass $m_q$ so that the light pseudoscalar meson is in the mass range 600--700~MeV. Over the $m_q$ range explored, $aM_\pi$ is roughly linear in $m_q$ for a given $\beta$.
\begin{figure}
    \centering
    \includegraphics[width=\columnwidth]{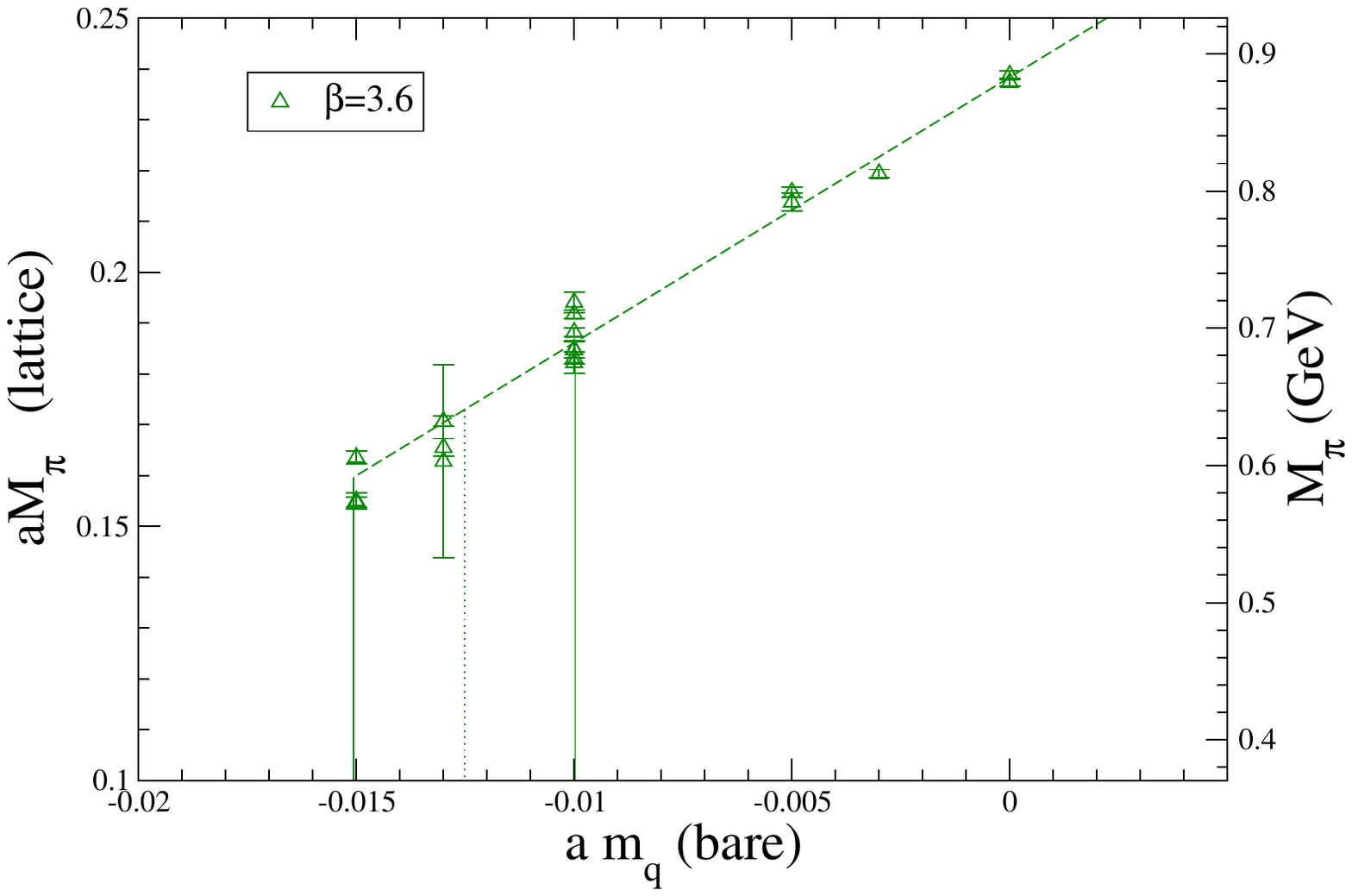}
    \caption{Tuning the light quark mass. The dashed diagonal line is a linear fit to ($m_q$,$aM_{\pi}$). The dashed vertical lines drop from the point the fit lines pass through $a(650~{\rm MeV})$. The vertical solid lines correspond to 600~MeV and 700~MeV. 
    \label{fig:mq_tune}}
\end{figure}
We generated 63 short tuning ensembles (500--600 Monte Carlo trajectories), iteratively refining our approach to the target points for three different $\beta$ values, with the intent of extending this calculation to include a continuum limit in the future.
For $\beta=3.6$, our ensemble at is $am_q=-0.013$, $am_c=0.25$ is very close to the target point and corresponds to $M_\pi=612(90)$MeV (with the uncertainty dominated by the lattice spacing determination).  Wilson-type fermion discretizations induce an additive renormalization of the quark masses, so bare light quark masses may be negative. We generated three ensembles, each with $L_t=64$ lattice units in the time direction, but varying spatial volumes of $L_s^3=32^3, 40^3, 48^3$. These correspond  roughly to spatial sizes of 1.6, 2.1 and 2.6 fm, respectively.

{\it Analysis and results.}---For each target ensemble we measure hadron correlators between operators corresponding to the [6] and the [15] irreps, in Eq.~\eqref{eqn:6 contraction}.  Additionally, we calculate pion and $D$ meson correlators. In each of these cases we smear the quark source operators and contract with both point and smeared sink operators. We used 32 to 128 sources per configuration. Measurement is on every tenth trajectory after 300 thermalization trajectories.

\begin{figure}
    \centering
    \includegraphics[width=0.9\columnwidth]{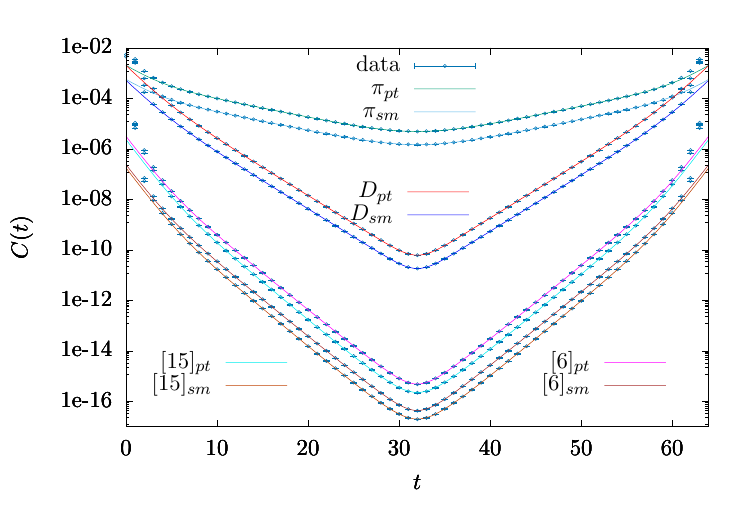}
    \caption{Eight hadron correlators used for simultaneous 2-state fits with $t_{\rm min}=5$ (see text). From top to bottom, the correlators represent: $\pi_{\rm pt}$, $\pi_{\rm sm}$, $D_{\rm pt}$, $D_{\rm sm}$, $[6]_{\rm pt}$, $[15]_{\rm pt}$, $[6]_{\rm sm}$, and $[15]_{\rm sm}$, with the subscripts ``pt" and ``sm" referring to point and smeared sink operators. Lattice data error bars are too small to be seen.}
    \label{fig:all_corrs}
\end{figure}

We then fit the resulting eight ensemble-averaged correlators simultaneously. The fit function contains 
\begin{equation}
    C_{P,s}(t) =\sum_{j=0}^{(N-1)}
    A_{P,s,j}\cosh\left(M_{P,j}(t-L_t/2)\right)
\end{equation}
for $P=\{\pi, D\}$ and 
\begin{multline}\label{eq:fitform}
    C_{P,s}(t) =
    B_{s}\cosh\left((M_D-M_\pi)(t-L_t/2)\right)\\
    +A_{P,s,0}\cosh\left((\Delta_{M_P}+M_{D,0}+M_{\pi,0})(t-L_t/2)\right)\\
    + \sum_{j=1}^{(N-1)} A_{P,s,j}\cosh\left(M_{p,j}(t-L_t/2)\right),
\end{multline}
for $p=\{ [6], [15]\}$, and $s=$\{point, smeared\} sink operators.
The overall fit form is
\begin{equation}
{\mathcal C}(t,P,s) = \delta_{P,P'}\delta_{s,s'}C_{P',s'}(t)
\end{equation}
The first term in Eq.~\eqref{eq:fitform} arises from the [6] and [15] operators also coupling to a combination of a $\pi$ propagating forward and a $D$ propagating backwards in time (and vice versa). This artifact of the periodic lattice must be accounted for, as it is lower in mass than the relevant [6] or [15] states. Correlators differing only by the sink smearing $s$ share the same mass. Figure~\ref{fig:all_corrs} shows our LQCD data for all these correlators as well as our resulting fits to these correlators for the $N=2$ case.

With these coupled fits we can directly extract the ground state mass shifts
\begin{eqnarray}
    \Delta M_{[6]}&\equiv& M_{[6]}-\left(M_D+M_\pi\right), \nonumber\\
    \Delta M_{[15]}&\equiv& M_{[15]}-\left(M_D+M_\pi\right).
\end{eqnarray}
In our analysis we perform both ground state ($N=1$) and ground plus excited state ($N=2$) fits for each $C(t)$. The fit range is $\left[t_{\rm min},\ldots, L_t-t_{\rm  min}\right]$. 
We vary the fit range of the data; a lower $t_{\rm min}$ allows more excited state contributions which may not be accommodated by the fit form. We use binned jackknife resampling (bin size equal to 15) to estimate the statistical uncertainty for each fit.
\begin{figure}
    \centering
    \includegraphics[width=\columnwidth]{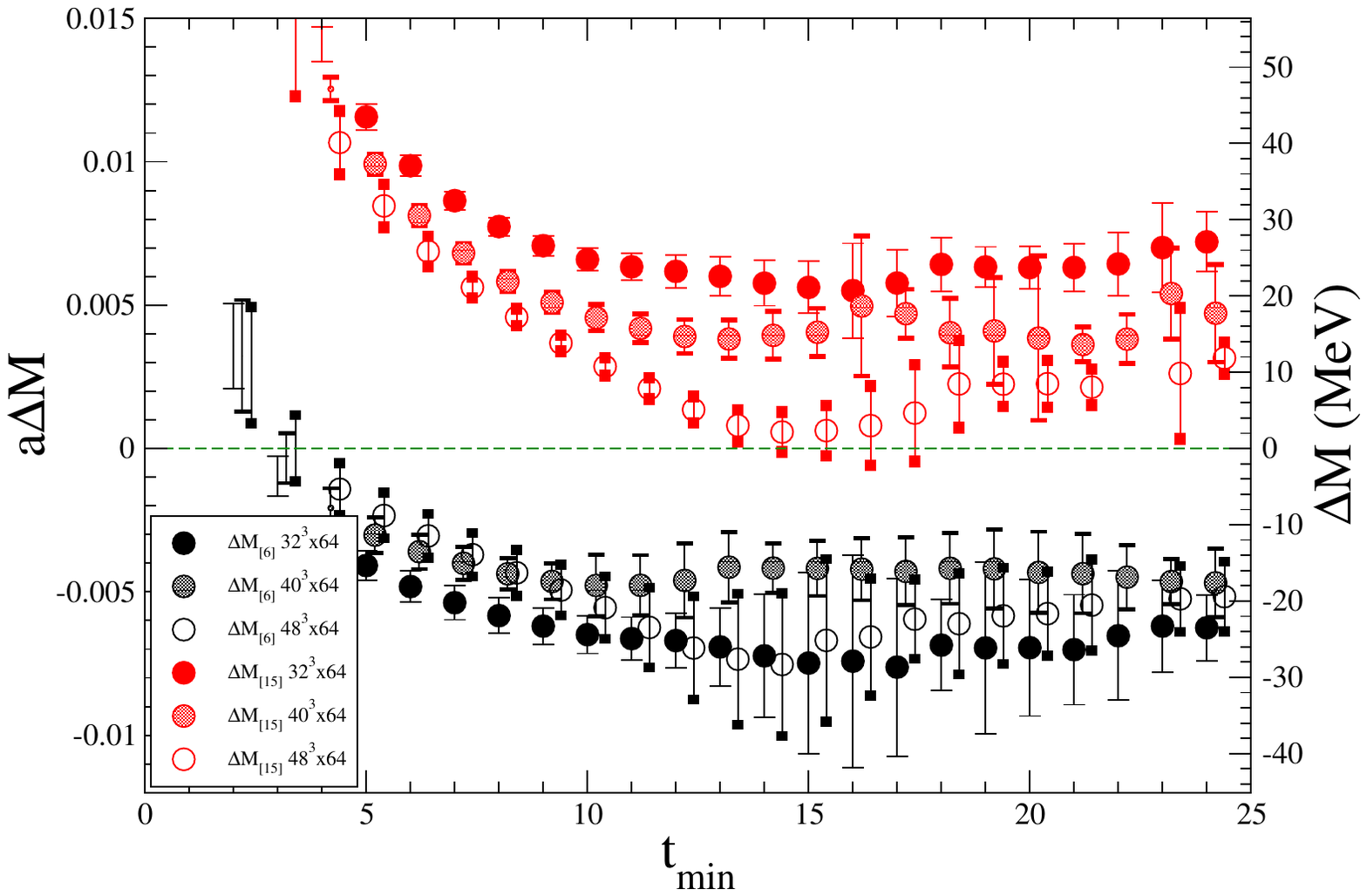}
    \caption{Mass shifts $\Delta{M_{[6]}}$ and $\Delta{M_{[15]}}$ from two-state fits  ($N=2$) as a function of $t_{\rm min}$. A visible symbol indicates that the fit confidence ($p$-value) is near 1. Error bars are larger on the $40^3$ fits because there are fewer measurements per configuration.}
    \label{fig:2state_fits}
\end{figure}

Figure~\ref{fig:2state_fits} shows the extracted mass shifts from 2-state fits. We also tried one state fits but there are residual $t_{\rm min}$-dependence and poor p-values for $t_{\rm min} < 13$. Two-state fits, by contrast, display very little $t_{\rm min}$-dependence, and achieve reasonable confidence by $t_{\rm min}\approx 4$. This indicates reasonable accounting of the excited states. 

{\it Discussion and Summary.}---We find that $\Delta M_{[6]}<0$ well in excess of statistical uncertainties. For $t_{\rm min}>10$ the fitted $\Delta M_{[6]}$ values show little or no discernible volume dependence. Taken together this {suggests} that the [6] is a bound state at this lattice spacing, {though more statistics and a detailed L\"uscher analysis would be required to resolve the volume dependence and truly identify the state as either bound or virtual.  But since} the goal of this work is to use LQCD to establish the non-perturbative behavior of the [6] and to investigate the interaction in the [15], we have not performed a continuum extrapolation in the lattice spacing {nor a L\"uscher analysis} as would be required for a precision calculation of the energy shifts {and its scattering parameters}.  As it stands, it looks likely that the binding energy of the [6] is in the range of $-10$ to $-30$~MeV, though we acknowledge that there are discretization effects that must be ascertained, such as those seen in, e.g., Ref.~\cite{Green:2021qol}. We are currently generating more ensembles at different volumes and lattice spacings to address this issue.

On the other hand, $\Delta M_{[15]}>0$ in every case, decreasing in value as the volume becomes larger. This is consistent with the [15] being a repulsive scattering state. Missing from this analysis is the [$\bar 3$] state, which is expected to be even more bound than the [6] state.  This state is additionally complicated due to the presence of disconnected diagrams. However, there are already lattice QCD
studies that demonstrate this state as the most strongly bound one with increasing binding for growing quark masses~\cite{Moir:2016srx,Gayer:2021xzv}, which allowed us to postpone the investigation of the [$\bar 3$] state
to a later publication.

The results of the lattice QCD study presented here are
exactly in line with the predictions for the non-perturbative
meson-meson dynamics derived from unitarised chiral perturbation
theory. {The negative energy shift in the [6] state establishes beyond doubt a pole close to threshold.  This is most probably a bound state but given the
current statistics a virtual state cannot be excluded.
In either case our results represent} the first  \emph{ab initio}
study showing non-trivial evidence for the molecular nature of the lightest non-strange charmed scalar mesons.
At the same time our findings are in clear conflict to
expectations from other models for the lightest singly heavy
scalar mesons: The existence of {this near-threshold pole} in the
flavor $[6]$ demonstrates unambiguously a multiquark structure
of the light scalars thus ruling out the applicability of the
traditional quark models for theses states. Moreover, the
repulsion observed in the flavor $[15]$ is in conflict with 
expectations from the tetraquark model
and explicitly excludes its realization in Ref.~\cite{Dmitrasinovic:2005gc}.
In this sense the lattice study presented here provides
important progress towards an understanding of the 
hadron spectrum.

{\it Acknowledgments.} We thank Alessandro Pilloni,  Evan Berkowitz and Andrea Shindler for useful discussions. We also thank David Wilson for his critical reading of our manuscript and our ensuing discussions.
TL is indebted to Andr\'{e} Walker-Loud for his insightful discussions related to backward propagating states and for politely correcting TL's ignorance in various subject matters. We are grateful to Balint Joo for assistance with CHROMA and to Kate Clark and Mathias Wagner for guidance with QUDA.

The authors gratefully acknowledge the computing time granted by the JARA Vergabegremium and provided on the JARA Partition part of the supercomputer JURECA\cite{jureca} at Forschungszentrum J\"{u}lich, through project CJIAS0421, and the Gauss Centre for Supercomputing e.V. (www.gauss-centre.eu) for funding this project by providing computing time on the GCS Supercomputer JUWELS\cite{juwels} at J\"{u}lich Supercomputing Centre (JSC), through project HADRONSEXTCONT.

This work is supported in part by the National Natural Science Foundation of China (NSFC) under Grant No.~11835015,
No.~12047503, and No.~11961141012, by the NSFC and the Deutsche Forschungsgemeinschaft (DFG, German Research
Foundation) through the funds provided to the Sino-German Collaborative
Research Center ``Symmetries and the Emergence of Structure in QCD''
(NSFC Grant No.~12070131001, DFG Project-ID 196253076 -- TRR110), and by the Chinese Academy of Sciences (CAS) under Grant No.~XDB34030000
and No.~QYZDB-SSW-SYS013.

\bibliographystyle{apsrev}
\bibliography{dpi}

\end{document}